\documentclass[twocolumn,aps,prb,superscriptaddress]{revtex4-1}
\usepackage[T1]{fontenc}\usepackage[latin1]{inputenc}
\usepackage{dcolumn,graphicx,color,booktabs,microtype,afterpage} 
\usepackage{sidecap}
\usepackage{times}
\usepackage[colorlinks,plainpages=false,linkcolor=blue,urlcolor=blue,citecolor=blue,pdfpagemode=UseNone,pdfstartview=FitBH]{hyperref}

\usepackage{amssymb}
\usepackage{amsmath}
\usepackage{graphicx}

\newcommand{\bcpo}{BaCo$_2$(PO$_4$)$_2$}
\newcommand{\bcao}{BaCo$_2$(AsO$_4$)$_2$}
\newcommand{\gbcpo}{$\gamma$-BCPO}
\newcommand{\J}{$J_1$-$J_2$-$J_3$}

\newcommand{\Seff}{ $S_{\text{eff}}$}
\begin{document}
%
%
%
%
\title
{Short range order in the quantum XXZ honeycomb lattice material BaCo$_2$(PO$_4$)$_2$}
\author{Harikrishnan S. Nair}
\email{hsnair@colostate.edu}
\affiliation{Department of Physics, Colorado State University, 200 W. Lake St., Fort Collins, CO 80523-1875, USA}
\altaffiliation{Current address:  Department of Physics, University of Texas El Paso, 500 W. University Ave., El Paso, TX 79968, USA}
\author{J. M. Brown}
\affiliation{Department of Physics, Colorado State University, 200 W. Lake St., Fort Collins, CO 80523-1875, USA}
\author{E. Coldren}
\affiliation{Department of Physics, Colorado State University, 200 W. Lake St., Fort Collins, CO 80523-1875, USA}
\author{G. Hester}
\affiliation{Department of Physics, Colorado State University, 200 W. Lake St., Fort Collins, CO 80523-1875, USA}
\author{M. P. Gelfand}
\affiliation{Department of Physics, Colorado State University, 200 W. Lake St., Fort Collins, CO 80523-1875, USA}
\author{A. Podlesnyak}
\affiliation{Neutron Scattering Division, Oak Ridge National Laboratory, Oak Ridge, TN 37831, USA}
\author{Q. Huang}
\affiliation{National Institute of Standards and Technology, Gaithersburg MD 20899, USA}
\author{K. A. Ross}
\email{Kate.Ross@colostate.edu}
\affiliation{Department of Physics, Colorado State University, 200 W. Lake St., Fort Collins, CO 80523-1875, USA}
\affiliation{Quantum Materials Program, Canadian Institute for Advanced Research (CIFAR), Toronto, Ontario M5G 1Z8, Canada}
\date{\today}
%
%
%
%
\begin{abstract}
We present observations of highly frustrated quasi two-dimensional (2D) magnetic correlations in the honeycomb lattice layers of the \Seff = 1/2 compound $\gamma$-\bcpo\ (\gbcpo).  Specific heat shows a broad peak comprised of two weak kink features at $T_{N1} \sim$ 6 K and $T_{N2} \sim$ 3.5 K, the relative weights of which can be modified by sample annealing.  Neutron powder diffraction measurements reveal short range quasi-2D order that is established below $T_{N1}$ and $T_{N2}$, at which two separate, incompatible, short range magnetic orders onset: commensurate antiferromagnetic correlations with correlation length $\xi_c = 60\pm2$ \AA  \ ($T_{N1}$) and in quasi-2D helical domains with $\xi_h = 350 \pm 11$ \AA \ ($T_{N2}$).  The ac magnetic susceptibility response lacks frequency dependence, ruling out spin freezing.   Inelastic neutron scattering data on \gbcpo\ is compared with linear spin wave theory, and two separate parameter regions of the XXZ \J\ model with ferromagnetic nearest-neighbor exchange $J_1$ are favored, both near regions of high classical degeneracy.   High energy coherent excitations ($\sim 10$ meV) persist up to at least 40 K, suggesting strong in-plane correlations persist above $T_N$.   These data show that \gbcpo\ is a rare highly frustrated, quasi-2D \Seff = 1/2 honeycomb lattice material which resists long range magnetic order and spin freezing.  

\end{abstract}
\maketitle
%
%
%
%

\section{Introduction}

\indent
Magnetic ground states and excitations of frustrated honeycomb lattices have been an active area of research, especially in connection with the search for quantum phases of matter such as quantum spin liquids (QSLs). A recent thrust in this direction is the study of the Kitaev anisotropic exchange Hamiltonian, which hosts QSL ground states \cite{kitaev2006anyons,trebst2017kitaev}.  The XXZ honeycomb model with competing $J_1$ (nearest neighbor), $J_2$ (second neighbor), and $J_3$ (third neighbor) interactions is also of significant interest, particularly for parameters that produce classical degeneracies \cite{fouet2001investigation, rehn2015classical}.   When quantum fluctuations are included, such classical degeneracies may favor disordered quantum phases, several of which have been proposed \cite{fouet2001investigation,cabra2011quantum, zhang2013exotic, zhang2014quantum, plekhanov2017emergent}.   

 Although many known honeycomb materials show evidence of competing interactions  \cite{viciu2007structure, dejongh2012magnetic, roudebush2013structure, seibel2013structure, zvereva2015zigzag,wong2016zig, wildes2017magnetic}, most of these appear to be far enough away from classical phase boundaries that they form long range ordered (LRO) states.  Interesting exceptions are Bi$_3$Mn$_4$O$_{12}$(NO$_3$) \cite{matsuda2010disordered}, InCu$_{2/3}$V$_{1/3}$O$_{3}$  \cite{moller2008structural}, and Na$_2$Co$_2$TeO$_6$ \cite{lefranccois2016magnetic, bera2017zigzag} which remain short range correlated well below their mean interaction strengths; however, for these materials there is also no consensus regarding the effective spin Hamiltonians which could account for their behavior.  Meanwhile, some Honeycomb-lattice based materials with significant static or dynamic structural disorder, namely Ba$_3$CuSb$_2$O$_9$ \cite{nakatsuji2012spin,katayama2015absence} and 6HB-Ba$_3$NiSb$_2$O$_9$ \cite{cheng2011high,darie2016new,quilliam2016gapless}, show QSL-like signatures, including the absence of LRO without spin freezing, and have been recently discussed in the context of a random-bond singlet phase \cite{uematsu2017randomness}.
\begin{figure}[!t]
\includegraphics[scale=0.4]{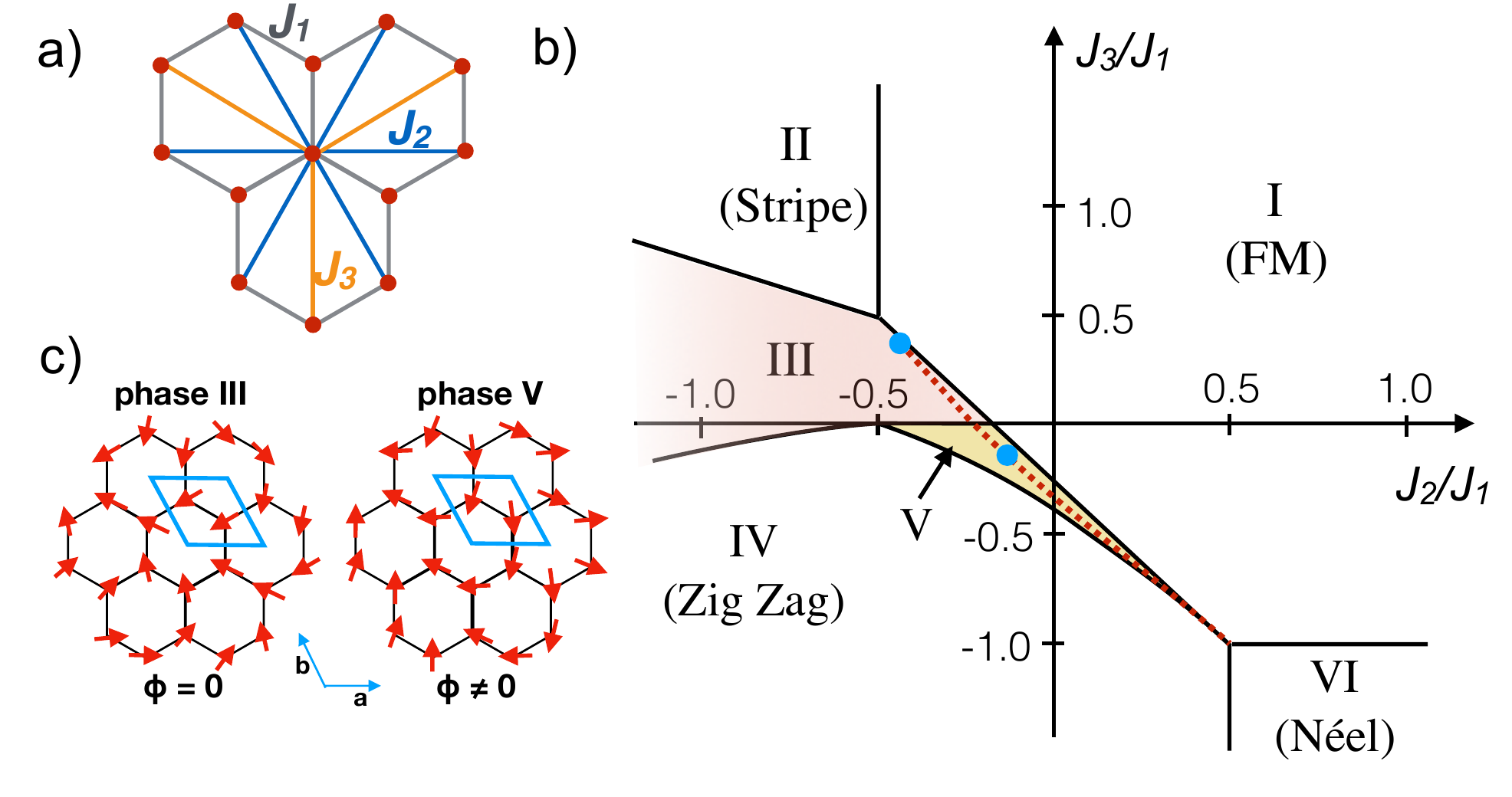}
\caption{(color online) a) Honeycomb lattice with interactions indicated.  b) Classical phase diagram for the $J_1$-$J_2$-$J_3$ XXZ honeycomb model with FM $J_1$ (< 0), reproduced from Ref. \onlinecite{fouet2001investigation}. Shaded regions (phases III and V) are helical phases.  The dotted red line indicates magnetic order with the helical ordering wavevector observed in \gbcpo, $|\vec{k}_h|$ = 0.25 r.l.u., and blue circles indicate possible locations for \gbcpo\ based on inelastic neutron scattering data. c) Representation of the helical order in Phase III and Phase V for parameter sets which closely match the inelastic neutron scattering data for \gbcpo.  The main difference between the two phases is the direction of the propagation vector ($\vec{k}_h = (0.146, 0.146, 0)$ in phase III vs. $\vec{k}_h = (0.25, 0, 0)$ in phase V) and the angle $\phi$ between the two spins in the primitive cell ($\phi = 0$ in phase III vs. $\phi \ne 0$ in phase V).} 
\label{fig:diagram}
\end{figure}

  Here we report on a short range correlated \Seff = 1/2  honeycomb lattice material, $\gamma$-\bcpo\ (\gbcpo) \cite{david2013puzzling}, for which the exchange parameters can be estimated due to the presence of helical short range order. We show that \gbcpo\ is positioned near regions in the \J\ model's parameter space with high classical degeneracy and phase competition, which may indicate that \gbcpo\ is a rare honeycomb lattice material that is proximal to a quantum disordered phase such as a QSL.
    
The \J\ XXZ honeycomb model takes the form,
\[ 
 H = \sum_{n=1}^{3}J_n\sum_{\langle i,j \rangle_n} (S_{xi}S_{xj} + S_{yi}S_{yj} + \lambda S_{zi}S_{zj}),
 \]
 where $i$ and $j$ run over the appropriate neighbor pairs (as shown in Figure \ref{fig:diagram}a), and $\lambda \in [0,1]$.   At the classical level this model hosts six ordered phases \cite{fouet2001investigation, rastelli1979non}, including four collinear phases (N\'eel, zig-zag, ferromagnet, and stripe) and two helical phases (phase III and V).  There is a well-known symmetry linking the phase diagrams for antiferromagnetic (AFM, $J_1 > 0$) and ferromagnetic (FM, $J_1<0$) nearest neighbor interactions \cite{fouet2001investigation}.  The two diagrams are mirror images across the $J_2 = 0$ line, with a relative 180$^{\circ}$ rotation of the moments on the two atoms of the Bravais lattice basis.   The classical phase diagram for $J_1<0$, which we show here is relevant to \gbcpo, is shown in Fig.~\ref{fig:diagram}b).   This phase diagram was explored years ago by Rastelli \emph{et al} \cite{rastelli1979non}, who found analytical solutions for the ordering wavevectors for helical phases III and V  \footnote{The analytical solution for phase III in Ref. \onlinecite{rastelli1979non} appears to have been misprinted and does not produce real values for the ordering wavevector throughout all of phase III; we have recalculated the ordering wavevectors using the Luttinger-Tisza method in this region (see Appendix \ref{sec:LT}).}.

Both $\gamma$-BCPO (space group $R\bar{3}$ with room temperature lattice parameters $a = 4.8554$ \AA, $c = 23.2156$ \AA) \cite{bircsak1998barium} and the isostructural \bcao\ (BCAO), were previously studied in the context of 2D XY models \cite{dejongh2012magnetic}.  These materials host undistorted magnetic honeycomb layers which are well-separated (7.9 \AA) by non-magnetic atoms.   BCAO can be made as large single crystals, amenable to detailed analysis by inelastic neutron scattering (INS) and directionally dependent magnetic susceptibility measurements, while the $\gamma$ phase of BCPO is metastable \cite{bircsak1998barium, david2013puzzling} and has so far only been made as small single crystals or powder samples.  Thus, BCAO has received the most attention.  It is known to be an example of an $S_\mathrm{eff}$ = 1/2 XY-like honeycomb lattice model, with FM nearest neighbor interactions, and it orders into an incommensurate magnetic phase with propagation vector $\vec{k} = (0.261, 0, -1.33)$ \cite{regnault1977magnetic, regnault2018polarized}.  The N\'eel transition, which is signaled by a sharp specific heat anomaly, appears to be preceded by a regime of Kosterlitz-Thouless behavior \cite{dejongh2012magnetic}.  However, BCAO's magnetic excitations do not conform to the expectations for a simple helical magnetic structure, conspicuously lacking a dispersion minimum at the ordering wavevector and instead displaying a gapped spectrum ($\Delta = 1.45$ meV) with a minimum at $Q$=0.  Recently, detailed spherical neutron polarimetry studies have shown that the magnetic structure and the resulting ferromagnetic fluctuations of BCAO result from weakly correlated ferromagnetic chains rather than a helical structure \cite{regnault2006investigation, regnault2018polarized}. 

Due to their structural similarity, $\gamma$-BCPO is expected to be magnetically similar to BCAO.  However, early reports noticed striking discrepancies in thermodynamic properties and type of magnetic correlations \cite{dejongh2012magnetic}.  Here we report on the unusual behavior of this material, presenting detailed thermodynamic, X-ray and neutron scattering experiments.  We show that \gbcpo\ is strongly frustrated, and in contrast to BCAO, remains short range correlated down to the lowest measured temperatures while tending towards two incompatible, yet coexisting, magnetic orders.  Further, through INS measurements, we have ruled out the gapped FM spin waves seen in BCAO, instead observing gapless modes consistent with helical short range order.  We have determined two candidate regions in the \J\ XXZ model parameter space consistent with the observed helical wavevector, both of which are proximal to classically degenerate regions of the phase diagram (including phase boundaries).  This suggests that the two incompatible ordering wavevectors are observed due to slight structural inhomogeneities favoring one state over the other in different regions of the sample, which is a signature of strong frustration.

\begin{figure}[!t]
\includegraphics[scale = 0.35]{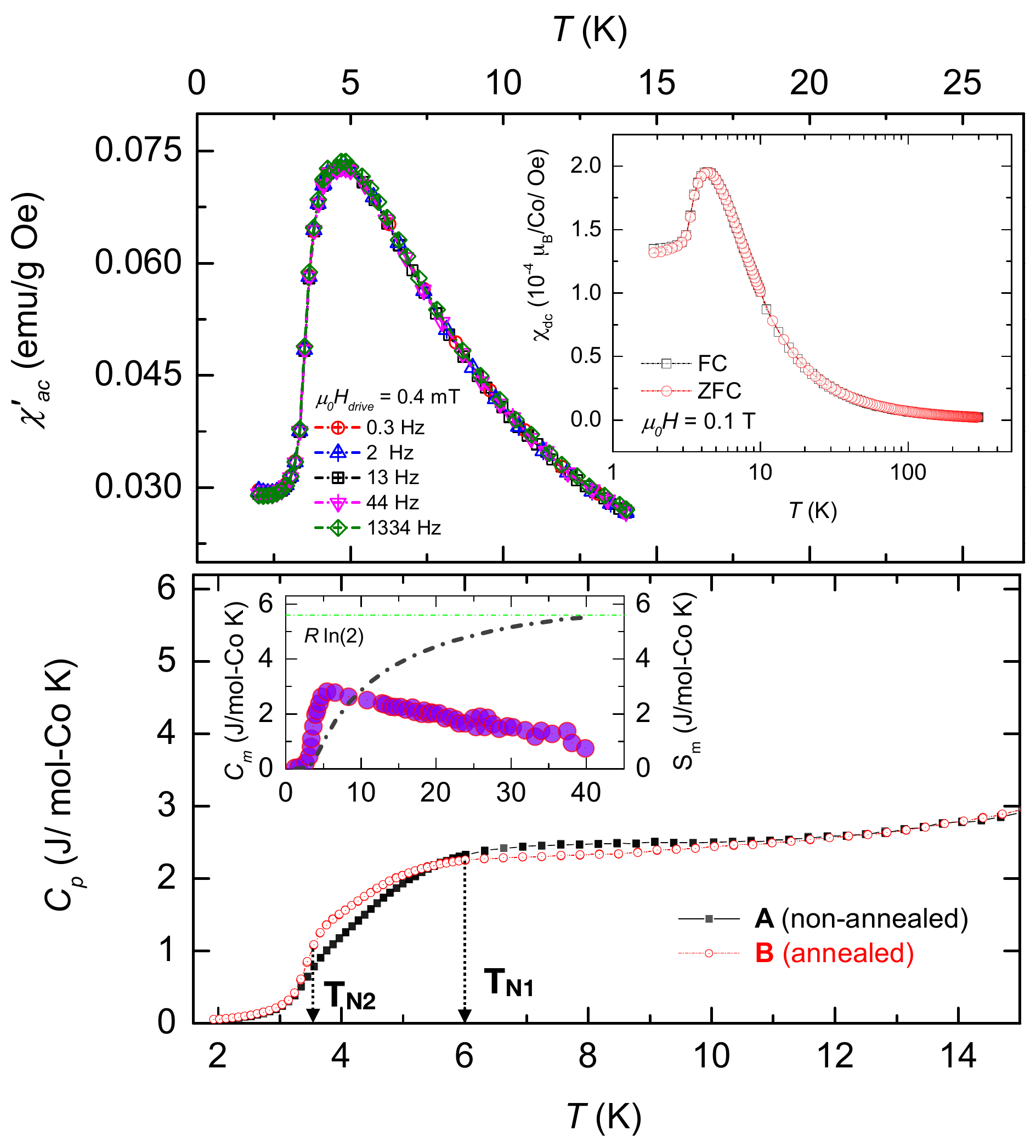}
\caption{(color online) Top: The real part of ac susceptibility of \gbcpo, which does not show any frequency dependence between 0.3Hz and 1334 Hz. Top inset: dc susceptibility under field cooled and zero field cooled conditions (no splitting observed). Bottom: Total specific heat of two samples of \gbcpo\ (A and B) which had different heat treatments, showing a slight change in weight of the weak features associated with $T_{N1}$ and $T_{N2}$ after annealing.   Inset: magnetic specific heat reproduced from Ref. \onlinecite{dejongh2012magnetic} and the entropy per Co$^{2+}$ derived from it. \label{fig_mag_cp}}
\end{figure}
\section{Experimental Method}
The polycrystalline samples of $\gamma$-BCPO used in the present study were prepared using a hydrothermal method following the procedure described in Ref. \onlinecite{bircsak1998barium} (Method 1), and a modified hydrothermal method (Method 2)  (Appendix \ref{sec:synth}).  Method 1 was used to produce crystals of approximate size $0.25 \times 0.25 \times 0.1$ mm$^3$, which were ground into a fine powder for thermodynamic measurements (sample A).  The powder was later annealed at 24~h at 100$^\circ$C with a heating rate of 1.2$^\circ$C/min (sample B) to investigate the effect of structural defects on the measurements.  Method 2 produced a fine powder of $\gamma$-BCPO with a 7.5 wt.\% impurity phase of Co$_2$(OH)(PO$_4$); 11.3 g of this powder was used for neutron scattering (sample C).  Although the impurity phase is known to be magnetic  \cite{rojo2002spin}, the transition temperature is very high (70 K) compared to relevant temperatures in \gbcpo\, and its magnetic signatures can be reliably removed from our data (Appendix \ref{sec:SXRD}).

Magnetic susceptibility was measured using a SQUID magnetometer (ac) and a vibrating sample magnetometer (dc) down to $T=$1.8 K (samples A and B).   Specific heat was measured down to 1.8~K using a thermal relaxation method (samples A and B).  Synchrotron x-ray diffraction (SXRD) patterns were recorded at $T=295$~K at beamline 11 BM ($\lambda$ = 0.41418~{\AA}) at the Advanced Photon Source, Argonne National Laboratory (samples A, B and C). Neutron powder diffraction (NPD) data were collected using the BT-1 32 detector neutron powder diffractometer ($\lambda$ = 2.0787(2) \AA, 60 minutes of arc collimation) at the NIST Center for Neutron Research (sample C, vanadium can) \cite{bt1}.  Time-of-flight inelastic neutron scattering (INS) experiments were performed at the Cold Neutron Chopper Spectrometer (CNCS) at Spallation Neutron Source, Oak Ridge National Laboratory (sample C, annular aluminum can).  INS data were collected for two incident neutron energies, $E_i$~=~3.07~meV and 14.9~meV in the ``intermediate'' chopper setting mode, producing energy resolutions of 0.06 meV and 0.45 meV at the elastic line, respectively \cite{ehlers2016cold}.  

\section{Results and Discussion}
\subsection{Thermodynamic measurements}
The dc magnetic susceptibility of $\gamma$-BCPO ($H = 0.1$ T) reveals a broad feature at $\approx$ 3~K, with no bifurcation between the zero field cooled (ZFC) and field cooled (FC) curves (Fig. \ref{fig_mag_cp}).  Comparable features are present in the ac susceptibility, $\chi_\mathrm{ac}'(T)$ at frequencies $f$ = 0.3 \textendash 1334~Hz (Fig~\ref{fig_mag_cp}, top).  No frequency dispersion is observed, ruling out a spin freezing transition. A broad anomaly in specific heat of \gbcpo \ is observed, centered at around 5~K (Fig~\ref{fig_mag_cp}, bottom).  The broad feature exhibits changes in slope (kinks) near $T_{N1} \sim 6 $ K and $T_{N2} \sim$ 3.5 K.  This is in stark contrast to the sharp $\lambda$-like phase transition reported for BCAO \cite{regnault1978specific}, and is consistent with short range order as we have confirmed by NPD (discussed below).  Further, the isolated magnetic contribution to the  specific heat, reproduced here from Ref. \onlinecite{dejongh2012magnetic}, indicates that spin correlations extend up to 40~K (Fig~\ref{fig_mag_cp}, bottom inset) at which temperature the entropy release reaches the total $R \ln 2$ expected for \Seff = 1/2 Co$^{2+}$ \cite{buyers1971excitations,goff1995exchange, abragam2012electron, ross2017single}.  The majority (78\%) of the entropy release occurs at temperatures above the broad peak in $C_p$.  We checked whether the lack of a transition to LRO could be due to lattice defects by comparing the specific heat of the same batch of \gbcpo\ before and after annealing (samples A and B respectively).   Sample B shows a slightly sharper heat capacity feature near $T_{N2}$, and a reduced weight near $T_{N1}$ but still lacks a conventional lambda anomaly.  SXRD data reveals that the structure of both samples is nearly identical, though it does indicate some lattice strain is relieved by annealing, while a small, unidentified impurity phase develops (Appendix \ref{sec:SXRD}).  None of our samples of \gbcpo\ show indications of stacking faults, which would manifest as asymmetric line-shapes in SXRD \cite{warren1941x}.  We investigated the temperature dependence of $C_p$ vs. $T$ below $T_{N1}$ and $T_{N2}$ and found that it does not follow any particular power law dependence over the limited temperature range available (1.8 K$<T<$3.5 K).  This is consistent with the findings summarized in Ref. \onlinecite{dejongh2012magnetic} that in \gbcpo\ and BCAO the specific heat does not follow a strictly $T^2$ temperature dependence, despite the otherwise well-established quasi-2D nature of the interactions in BCAO.

\begin{figure}[!t]
	\includegraphics[scale=0.5]{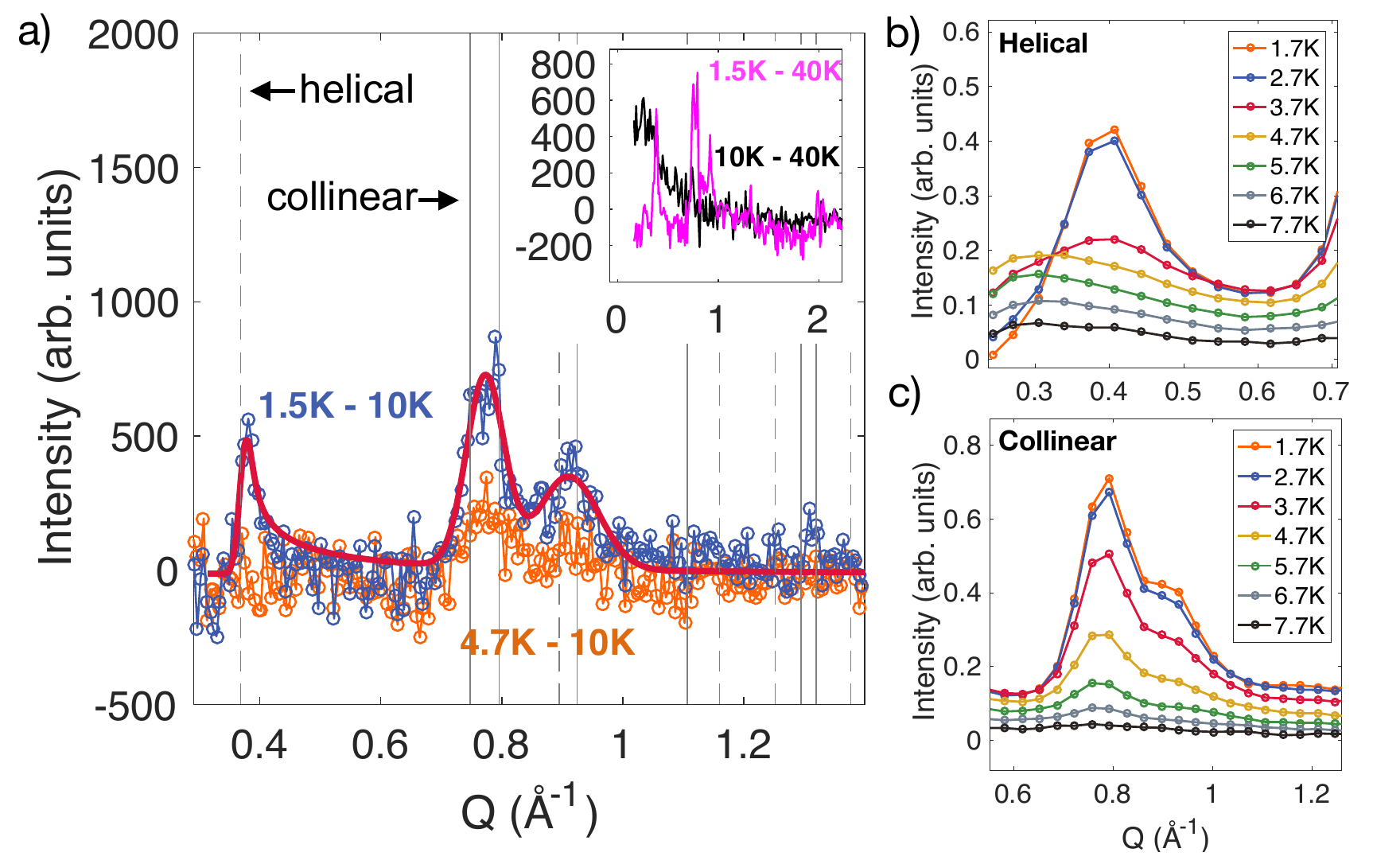}
	\caption{(color online) a) Neutron powder diffraction intensity (from BT-1) of \gbcpo\ at 1.5~K (blue) and 4.7~K (orange) after subtracting 10~K data with its diffuse background removed. The $Q$ values expected for long range helical and collinear AFM magnetic peaks are indicated by dashed and solid vertical lines, respectively. The solid red line is a fit (see main text). Inset: 1.5~K (magenta) and 10~K (black) data after subtracting the 40~K data, showing increased intensity near $Q = 0$ at 10~K, which indicates FM correlations.  This diffuse scattering is reorganized into broad magnetic peaks at 1.5~K.  b) and c) Detailed temperature dependence of the elastic magnetic scattering (from CNCS, after subtracting the elastic scattering at 10~K), showing different onset temperatures of the helical and collinear peaks.  Note that the $Q$ resolution at CNCS is more relaxed than at BT-1, obscuring the Warren lineshape for the lowest angle peak. \label{fig_bt1}}
\end{figure}

\begin{figure*}[!ht]
\includegraphics[scale = 0.5]{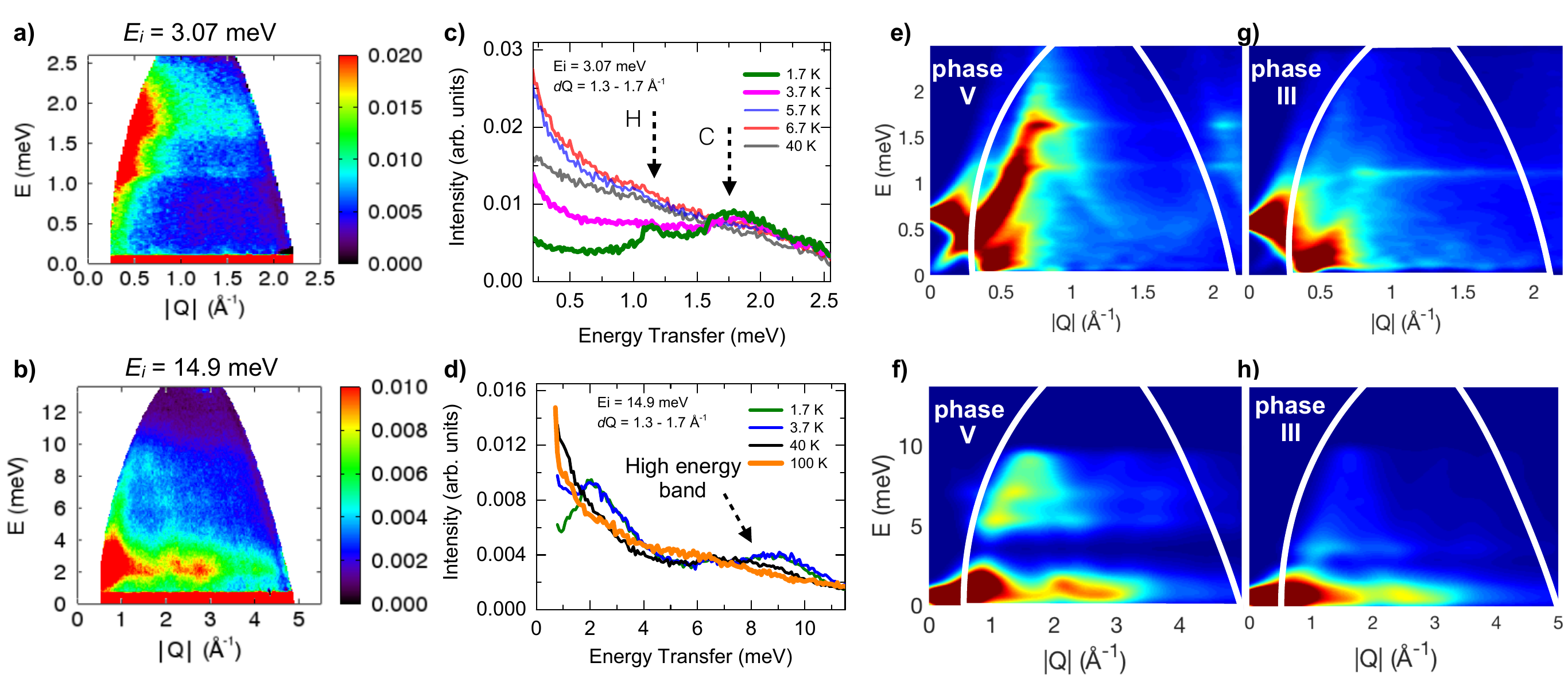}
\caption{ (color online) a) and b) Inelastic neutron scattering spectra at 1.7~K, with $E_i$ = 3.07 meV and $E_i$ = 14.9 meV, respectively.  c) and d) Constant $Q$ cuts (integrated from 1.3 to 1.7 \AA$^{-1}$) for $E_i$ = 3.07 meV and 14.9 meV, respectively, presented for different temperatures.  The contributions from helical (H) and collinear (C) ordering are indicated by arrows in c). e) and f) Calculated powder averaged spin excitations for parameter set 1 (see main text), shown for two different energy ranges.  g) and h) Calculated powder averaged spin excitations for parameter set 2. Additional INS data are shown in Appendix \ref{sec:INS}. \label{fig_cncs}}
\end{figure*}

\subsection{Neutron Powder Diffraction}
The static magnetic correlations in \gbcpo \ were investigated by neutron powder diffraction (NPD) on sample C.  In Figure ~\ref{fig_bt1} the intensity versus momentum transfer ($Q$) plot of the diffracted intensity at 1.5~K and 4.7~K are presented after subtraction of the data at 10~K ($T > T_{N2}$ and $T_{N1}$). Two separate ordering wavevectors are observed, as previously reported for \gbcpo\ \cite{dejongh2012magnetic}.  The magnetic peaks are approximately seven times broader than the instrument resolution ($dQ \sim 0.014$ \AA$^{-1}$ at $Q=0.4$ \AA$^{-1}$) \cite{bt1}.  The lowest $Q$ reflection (peaked at 0.38 \AA$^{-1}$), which onsets below $T_{N2}$, presents a clear Warren line shape, i.e., diffuse scattering intensity characteristic of 2D short-range order, where a sharp rise of intensity at low $Q$ and a slow fall towards high $Q$ is discernible.  The Warren shape indicates that this is a $(HK0)$ reflection arising from quasi-2D magnetic correlations in the honeycomb layers \cite{warren1941x}. A fit of the lowest angle reflection to the Warren line shape convoluted with the instrument resolution gives a planar correlation length of $350\pm11$ \AA\ and a central peak position in the $HK$ plane of $Q_0 = 0.373 \pm 0.001$ \AA$^{-1}$ (Appendix \ref{sec:NPD}).  The latter is consistent with helical ordering wavevectors of $\vec{k}_h$ = (0.25, 0, 0) or (0.146, 146, 0), each of which is relevant to different helical phases in the \J\ model, and both of which result in $Q_0$ = 0.374 \AA$^{-1}$. \footnote{Using the refined lattice parameters and instrument corrections obtained from the nuclear structure at 4.7K (Appendix \ref{sec:NPD}}    Two other broadened reflections are observed near 0.78 and 0.92 \AA$^{-1}$ (the lower of the two gives a correlation length of 60 $\pm$ 2 \AA), which are fit adequately with Gaussians. These higher $Q$ reflections onset below $T_{N1}$, gain intensity as the temperature is further lowered, and persist below $T_{N2}$.   No single ordering wavevector can account for all of the observed peak positions, but a combination of $\vec{k}_h$ (helical order) and $\vec{k}_c $ (collinear AFM, either stripe or zig-zag) wavevectors can, consistent with the original work on \gbcpo\ \cite{dejongh2012magnetic}.    The predicted positions of peaks arising from $\vec{k}_h$ and $\vec{k}_c$ are marked in Fig.~\ref{fig_bt1} by dashed and solid lines, respectively.  Note that the apparent central positions of broadened reflections arising from short range correlations are expected to appear at an offset in $Q$ compared to the corresponding long range ordered states, as observed here.

The coexistence of collinear and helical short range orders in \gbcpo, as well the absence of LRO, suggest that the material's effective spin Hamiltonian lies near a phase boundary in the classical phase diagram. A likely scenario is that the two magnetic orders arise from different spatial regions in the sample. The weight of the features in the heat capacity corresponding to $T_{N2}$ (onset of $\vec{k}_h$) and $T_{N1}$ (onset of $\vec{k}_c$) can be influenced by sample annealing,  suggesting that slight structural changes tip the balance between the different short range orders. This may also imply that bond disorder, induced by lattice disorder, could play a role in suppressing both LRO and spin freezing in BCPO, as suggested by recently-developed theories of the QSL-like random-singlet state \cite{uematsu2017randomness}.

The observed modulus of the short range helical wavevector ($|\vec{k}_h| = 0.25$ r.l.u. = 0.373~\AA$^{-1}$) constrains the exchange parameters for \gbcpo \ to two lines that pass through phase III ($\vec{k}_h = (0.146, 146, 0)$) and phase V ($\vec{k}_h = (0.25, 0,0)$), shown as red dotted lines in Fig.\ref{fig:diagram} b) (Appendix \ref{sec:LT}).   These lines approach borders with the stripe or zig-zag collinear antiferromagnet phases.  In phase III, the line approaches the highly degenerate point $J_2/J_1$ = -0.5, $J_3$/$J_1$ = 0.5 which borders on three phases; I (FM), II (stripe), and III (helical).  In phase V, the more negative $J_3$/$J_1$ becomes along the line, the nearer to the phase boundaries with zig-zag and FM orders the line is.  We now show that INS can be used to narrow down the parameters to more specific points in these two phases.

\subsection{Inelastic Neutron Scattering}
The magnetic excitations of \gbcpo \ measured by INS are presented in Figure~\ref{fig_cncs} for incident energies 3.07 meV (top row) and 14.9 meV (bottom row), along with representative calculations from linear spin wave theory (LSWT) \cite{Toth2015}.   Panels a) and b) show the intensity vs. $Q$ maps at the lowest measured temperature, $T=$1.7~K.  Three features are apparent: 1) intensity increases toward $Q = 0$ (but the spectrum is not gapped or peaked at $Q=0$, unlike BCAO, see Appendix \ref{sec:INS}) which is consistent with ferromagnetic nearest neighbor exchange, 2) referring to panel a) there are two ``flat band'' features, one at 1.2 meV and the other at 1.7 meV, indicating a high density of states for these energies, and 3) referring to the high incident energy scan in panel b), the two flat bands merge into a single intense band due to broader energy resolution, and a weaker high energy part of the dispersion is observed to extend up to $\sim$10 meV.  The temperature dependence of the latter two features is revealing (panels c and d). At $T=3.7$~K, between T$_{N1}$ and T$_{N2}$, the lowest flat band (1.2 meV) vanishes, while the other (1.7 meV) persists up to 5.7~K ($\sim T_{N1}$) (panel c).  This suggests that the former arises from the helical short range ordering, while the latter arises, at least in part, from the collinear short range ordering.  As the temperature increases above T$_{N1}\sim$ 6 K, the low energy intensity becomes more uniform and decreases monotonically as $Q$ increases, consistent with FM paramagnetic fluctuations (panel c and d).  However, the higher energy band around 10 meV survives up to 40~K (panel d).  We therefore suggest that this band corresponds to excitations within the 2D honeycomb layers, which arise from 2D correlations that extend up to 40~K, as also evidenced by the magnetic specific heat.

In Figure \ref{fig_cncs}, panels (e)-(h) show the powder averaged dynamic structure factor from LSWT, using two parameter sets in the \J\ XXZ honeycomb lattice model.  Directly fitting the INS data is made impractical due to the presence of additional excitations arising from the collinear AFM phase (namely the flat band near 1.7 meV) in addition to the likely strong quantum effects, as well as lack of LRO.  Yet, it is possible to narrow down the exchange parameters using two features that are compatible with the helical order.  The first feature is the flat band at 1.2 meV, and the second is the weak band of excitations extending up to 10 meV.   By calculating the powder averaged spectrum at many points along the $|\vec{k}_h| = 0.25$ r.l.u. lines for different choices of $\lambda$ between 0 (XY) and 1 (Heisenberg), we obtained two sets of parameters which adequately reproduce the features attributed to helical short range order, while not introducing any extraneous features.  We note that the presence of a higher energy band is ubiquitous along this line, however for some parameter regimes its intensity relative to the lower band is much too high, and these parameters were ruled out (Appendix \ref{sec:LSWT}).   Further, the relative energy of the top of this band compared to the lowest flat mode strongly constrains the parameters.  We find that we can reproduce the main features of the helical excitations for the following two sets of parameters ($J$'s in meV):
\begin{align}
 J_1&=-4.33 , J_2 = 0.54 , J_3 = 0.67, \lambda = 0.85,  \\ 
 J_1&=-4.27 , J_2 = 1.92 , J_3 = -1.75 , \lambda = 0.40.
 \end{align}
These sets of parameters are indicated by blue dots in Fig. \ref{fig:diagram}~b.   We emphasize here that the ``best'' parameters (with associated confidence intervals) cannot be explicitly stated, since we have not performed numerical fits to the data due to the limitations of this comparison, as discussed above. As an estimate of the range of validity, we note that these parameters can be allowed to vary individually by $\sim$ 8\% of the values stated above, and still produce qualitatively similar results.   For comparison, parameters for BCAO were suggested to be, based on INS data in a small applied field, $J_1$=-3.27 meV, $J_2$ = -0.112 meV, $J_3$ = 0.86 meV, $\lambda$ = 0.4  (phase V) \cite{dejongh2012magnetic}, although it should be noted that these parameters do not reproduce the zero field spin wave spectrum in BCAO \cite{dejongh2012magnetic,regnault2006investigation,regnault2018polarized}.  

\subsection{Discussion}
Both suggested spin Hamiltonians for \gbcpo \ carry interesting implications.  Set 1 (phase V) implies \gbcpo \ is Heisenberg-like with a helical ordering wavevector of $\vec{k}_h = (0.25, 0, 0)$.  Wavevectors where $\vec{k} = \vec{G}/2$ or $\vec{G}/4$, with $\vec{G}$ a reciprocal lattice vector, lead to continuous classical degeneracies in the Heisenberg model \cite{fouet2001investigation}.  Far from phase boundaries, ``order by disorder'' selects one of the many possible classically degenerate ordered states, e.g. the collinear zig-zag and stripe phases when $\vec{k} = \vec{G}/2$. However, at the boundary between phase III and V in the FM nearest neighbor model, evidence for a QSL appears \cite{fouet2001investigation}.  Set 1 suggests \gbcpo \ is quite close to this instability of magnetic order.  However, how far this putative QSL extends into phase V or III has not yet been investigated.    Meanwhile, parameter set 2 suggests \gbcpo \ is XY-like, and puts it near the FM ``maximally frustrated'' point in the phase diagram.  The AFM nearest neighbor analog of this point was recently shown to host a classical spin liquid, with additional nematic order in the XY model \cite{rehn2015classical}.  Such a highly degenerate point is a natural place for a QSL in the quantum model, though this has not been explored theoretically.  Given the possible proximity of \gbcpo \ to the FM nearest neighbor analog of this highly degenerate point, a theoretical study of the quantum model near this point would be of particular interest.  To further distinguish between parameter sets 1 and 2, a detailed study of the magnetic correlations in \gbcpo \ on the available (small) single crystal samples may be successful.  It would also be of great utility to determine the single-ion anisotropy in \gbcpo, which directly influences the value of $\lambda$ \cite{goff1995exchange}.  Finally, a recent manuscript suggests that Co$^{2+}$ honeycomb lattice materials may host Kitaev interactions \cite{liu2017pseudospin}.  The magnitude of the Kitaev term in \gbcpo\ and BCAO would be of great interest to determine.

\section{Conclusions}
We have shown that the \Seff = 1/2 honeycomb material \gbcpo\ displays competing short range magnetic orders which onset below $T_{N1} \sim 6$ K and $T_{N2} \sim 3.5$ K.  We establish here the material's proximity to classically degenerate regions in the \J\ XXZ phase diagram, providing a rationale for its stubborn resistance to forming long range magnetic order or a spin-frozen state. We hope that this study inspires further theoretical work on the often overlooked ferromagnetic nearest neighbor Honeycomb lattice model and its possible quantum disordered phases.

\begin{acknowledgements}
We acknowledge J. R. Neilson, A.E. Maughan,  J. Kurzman, and S. Folkman for assistance with sample preparation, and R. Moessner for helpful discussions.  We acknowledge the support of the National Institute of Standards and Technology, U.S. Department of Commerce, as well as Oak Ridge National Laboratory, U. S. Department of Energy, in providing the neutron research facilities used in this work. We acknowledge the support of Argonne National Laboratory, U.S. Department of Energy, in providing the synchrotron facility used in this work.  This research was supported by the National Science Foundation Agreement No. DMR-1611217.
 \end{acknowledgements}

\appendix
\section{Sample Synthesis}
\label{sec:synth}
We produced powders of barium cobalt phosphate (\gbcpo) by hydrothermal synthesis, using two methods.  

\begin{itemize}
\item Method 1:
We adapted the synthesis method reported in Ref. \onlinecite{bircsak1998barium} by adjusting relative proportions of reactants in an effort to produce larger crystals.

We combined H$_2$O (12 mL), BaCO$_3$ (0.310 g), CoBr$_2\cdot$3H$_2$O (0.860 g), guanidinium carbonate [C(NH$_2$)$_3$]$_2$CO$_3$ (0.6231g), and 85\% H$_3$PO$_4$  (0.732 mL). The reaction is successful if the reactants are added in the order listed.  The reactants were mixed with a stir rod in a 23 mL teflon-lined hydrothermal autoclave (producing a transparent, bright pink solution) and heated to 180$^\circ$C for 72hr. The autoclaves were then left to cool to ambient temperature over several hours. Hexagonal shaped crystals of \gbcpo\ (average size of approximately 0.25 x 0.25 x 0.1 mm$^3$, bright pink color) were recovered after vacuum filtering the content of the autoclave and left to dry in air, and were manually separated from other powdered precipitates (BaCoP$_2$O$_7$) and washed with water.  The obtained crystals were then ground into a fine powder.

\item Method 2:  We designed a new hydrothermal recipe, which we found produces fine powders of \gbcpo, to obtain a large quantity of powdered material for neutron scattering (sample C).

We combined H$_2$O (12 mL), BaCl$_2$ (0.713 g), CoCl$_2$ (1.390 g), and Na$_3$PO$_4$  (0.958 g).  The reactants were mixed with a stir rod in a 23 mL teflon lined hydrothermal autoclave (producing a transparent, bright pink solution) and heated to 180$^\circ$C for 72hr.  The autoclaves were then left to cool to ambient temperature over several hours.  Fine powders of \gbcpo\ (bright pink) with a $\sim$ 7\% impurity of Co$_2$(OH)(PO$_4$) were recovered after vacuum filtering the product of the reaction.

\end{itemize}

\section{Synchrotron X-ray Diffraction Results}
\label{sec:SXRD}

\begin{figure}[!h]
\includegraphics[scale=0.35]{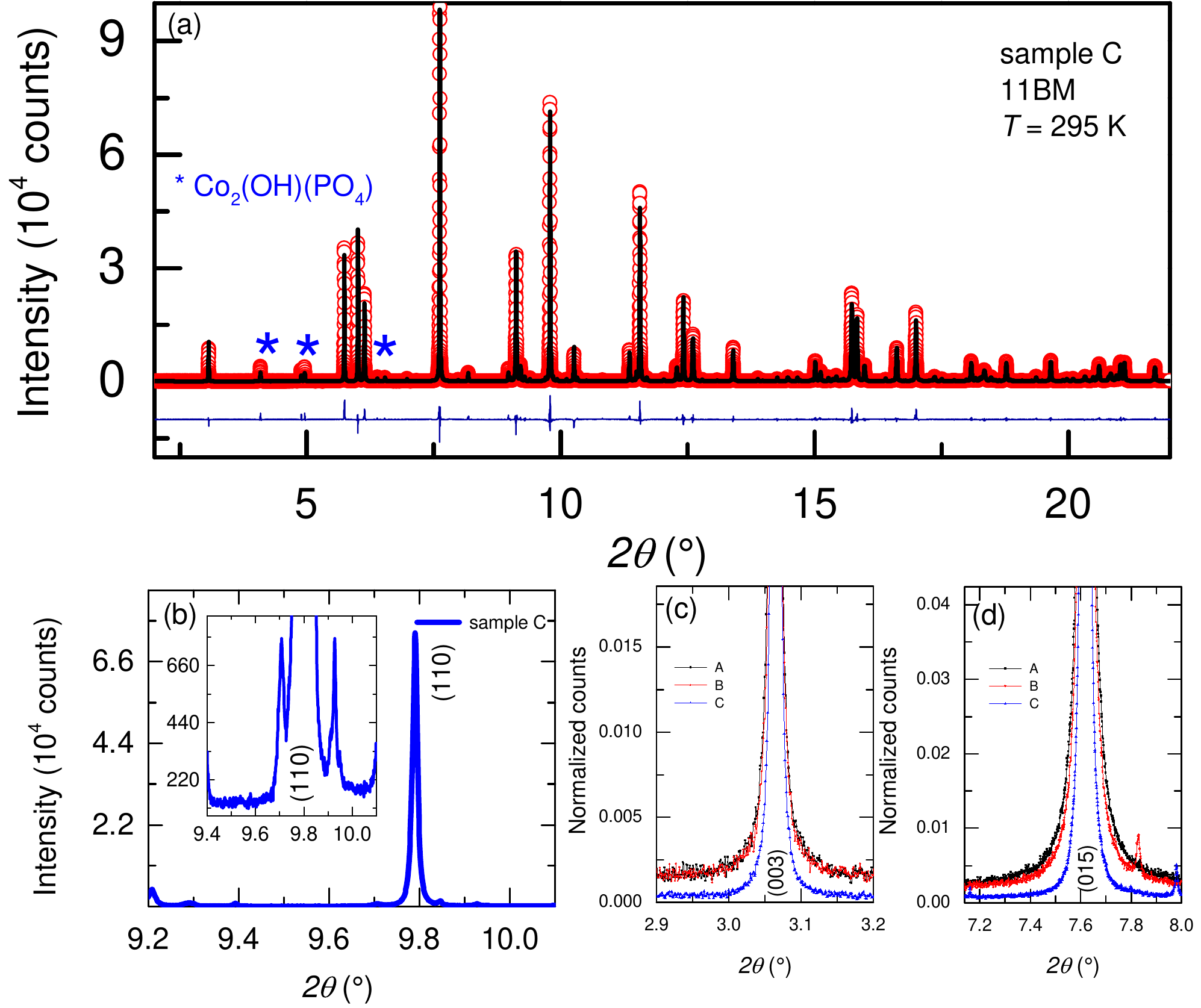}
\caption{a) SXRD results from Sample C (used for neutron scattering) at $T =$ 298 K.  A  7.5 wt.\% impurity phase of Co$_2$(OH)(PO$_4$) is present. b) Detailed view of an HK0 type reflection from sample C, showing the absence of a Warren lineshape which might have been expected to arise from stacking faults in the nuclear structure. c) Comparison of the shape of reflections for samples A, B, and C (normalized to the same peak height).   Widening at the base of the reflections is observed in sample A and B compared to C.  d) The broadening of some of the peaks is reduced after annealing (compare sample A to B).  
\label{fig:SXRD}}
\end{figure}

Figure \ref{fig:SXRD} shows results from the Synchrotron X-ray Diffraction (SXRD) experiment on three samples of \gbcpo.  All samples show good agreement with the published crystal structure of \gbcpo\ \cite{david2013puzzling}, with additional impurities of Co$_2$(OH)(PO$_4$) (Sample C) and an unidentified impurity (Sample B).   Sample C, which was used for neutron scattering, does not show any evidence for asymmetric Warren lineshapes of the structural Bragg peaks (panel b), ruling out structural stacking faults as a source for the quasi-2D magnetic correlations observed by NPD (Appendix \ref{sec:NPD}).  Sample A, produced by Method 1 (Appendix \ref{sec:synth}), was annealed as described in the main text to produce sample B.  Both samples A and B show an increased broadening at the base of the structural peaks compared to sample C (panels c and d), likely indicating increased levels of lattice strain.    The strain is partially relieved by annealing (panel d).  

\subsection{Magnetic properties of impurity phase}

The Co$_2$(OH)(PO$_4$) impurity found in samples made by Method 2 is known to display a magnetic ordering transition around 70~K and a possible spin freezing transition around 15~K \cite{rojo2002spin} but neither of these features correlate with the temperature dependence of our observations on \gbcpo, and neutron scattering responses are expected to be either very weak (inelastic) or are observed to be approximately constant in temperature below 40~K (elastic) and therefore can be removed using a high temperature subtraction.

\section{Neutron Powder Diffraction}
\label{sec:NPD}
\subsection{Short range correlations and Warren lineshape}

\begin{figure}[!h]
\includegraphics[scale=0.35]{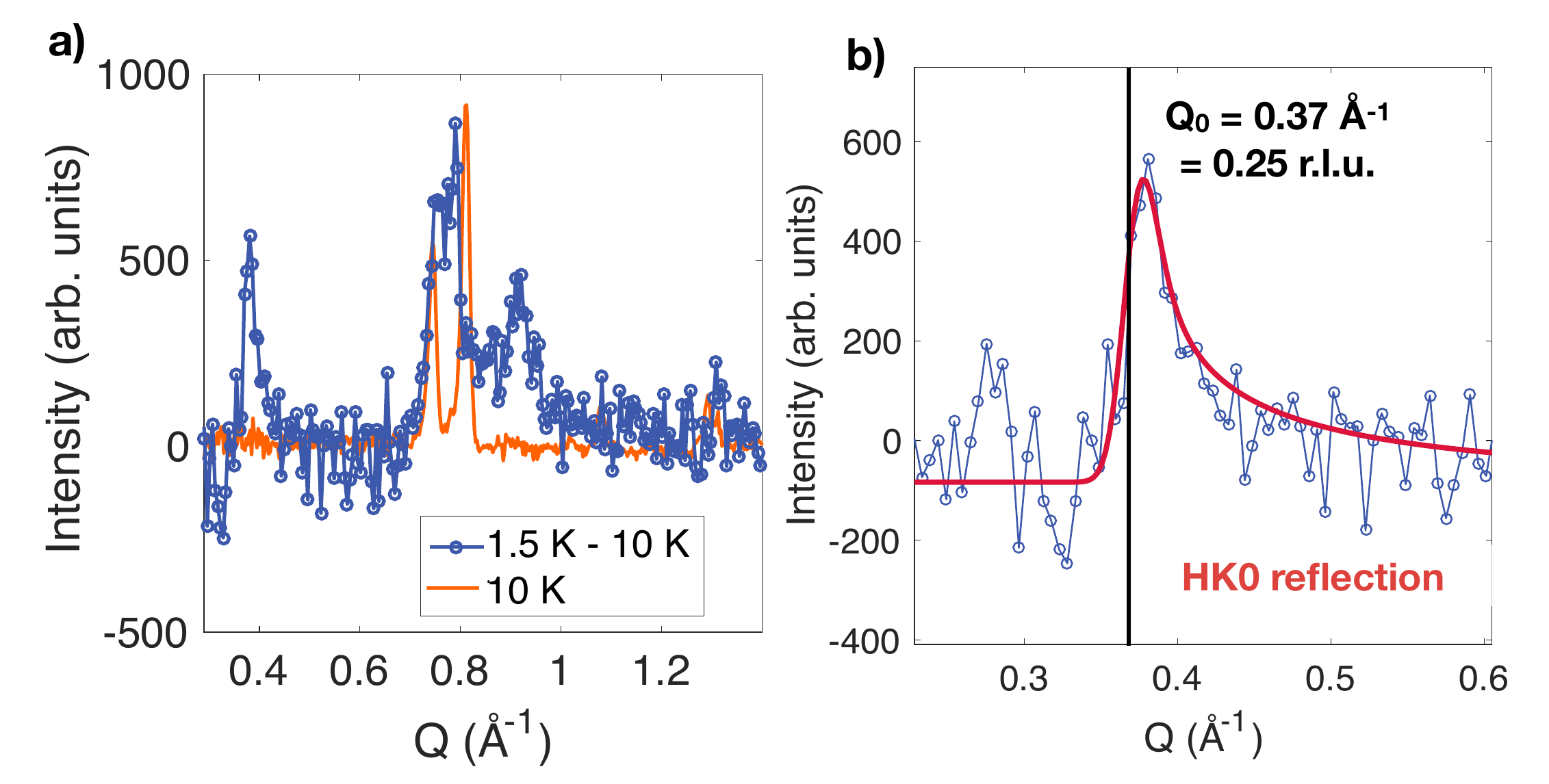}
\caption{a) Comparison of linewidths of nuclear (10 K) to magnetic (1.5 K - 10 K) reflections from neutron powder diffraction. The relative intensities of the two curves have been scaled to produce the same approximate peak height. b) Detailed view of the lowest angle reflection (1.5 K - 10 K) and fit to the warren function convolved with the instrument resolution.} 
\label{fig:BT1linewidth}
\end{figure}

The magnetic Bragg peaks shown in Figure 3 in the main text are significantly broader than the instrumental resolution.  This is illustrated in Figure \ref{fig:BT1linewidth} a, where the total scattering at 10 K, which is dominated by resolution-limited nuclear Bragg peaks, is overlaid by the 1.5 K - 10 K difference curve, which is purely magnetic scattering.  The broad magnetic reflections indicate short range magnetic correlations, rather than a long range ordered state, and appear below $T_{N1}$ and $T_{N2}$.

The lowest angle magnetic reflection displays a characteristic asymmetric shape known as a Warren lineshape, which arises from the constant $Q$ Ewald sphere cutting through a \emph{rod} of scattering arising from 2D correlations \cite{warren1941x,wills2000magnetic}.  We fit the lowest $Q$ reflection from \gbcpo\ using a convolution of a Gaussian instrumental resolution (with FWHM = 0.014\AA$^{-1}$) \cite{bt1} and the equation below (following Ref. \onlinecite{wills2000magnetic}):
\[
I(Q) = K f^2_mF(a)[2Q(\lambda/4\pi) + Q^{-1/2}(\lambda/4\pi)^{-2} -2]\bigg(\frac{QL}{4\pi^{3/2}}\bigg)^{1/2}\]
where,
\[ F(a) = \int_{0}^{10} \mathrm {exp}[-(x^2 - a)^2]dx,\]
with $a$ = $L(Q - Q_0)/2\sqrt{\pi}$, where $Q_0$ is the $Q$ value of the center of the scattering rod in the H-K plane (i.e., magnitude of the wavevector to the center of the reflection in the honeycomb layer reciprocal lattice).  $f_m$ is the magnetic form factor for Co$^{2+}$ \cite{brown2006magnetic}, $K$ is a scale factor, $\lambda$ is the incident wavelength of neutrons, and $L$ is the 2D spin-spin correlation length. The fit is shown in Fig.~\ref{fig:BT1linewidth} b), and we obtain the 2D spin-spin correlation length as $L = \xi_h = 350 \pm 11${\AA}, while $Q_0 = 0.373 \pm 0.001$\AA$^{-1}$.
The quasi-2D nature of the interactions and the resulting short-range spin correlations in \gbcpo\ are confirmed through the analysis presented here, as well as the identification of the lowest angle reflection as an HK0 peak.

\subsection{Nuclear Structure Refinement}
\begin{figure}[!h]
\includegraphics[scale=0.45]{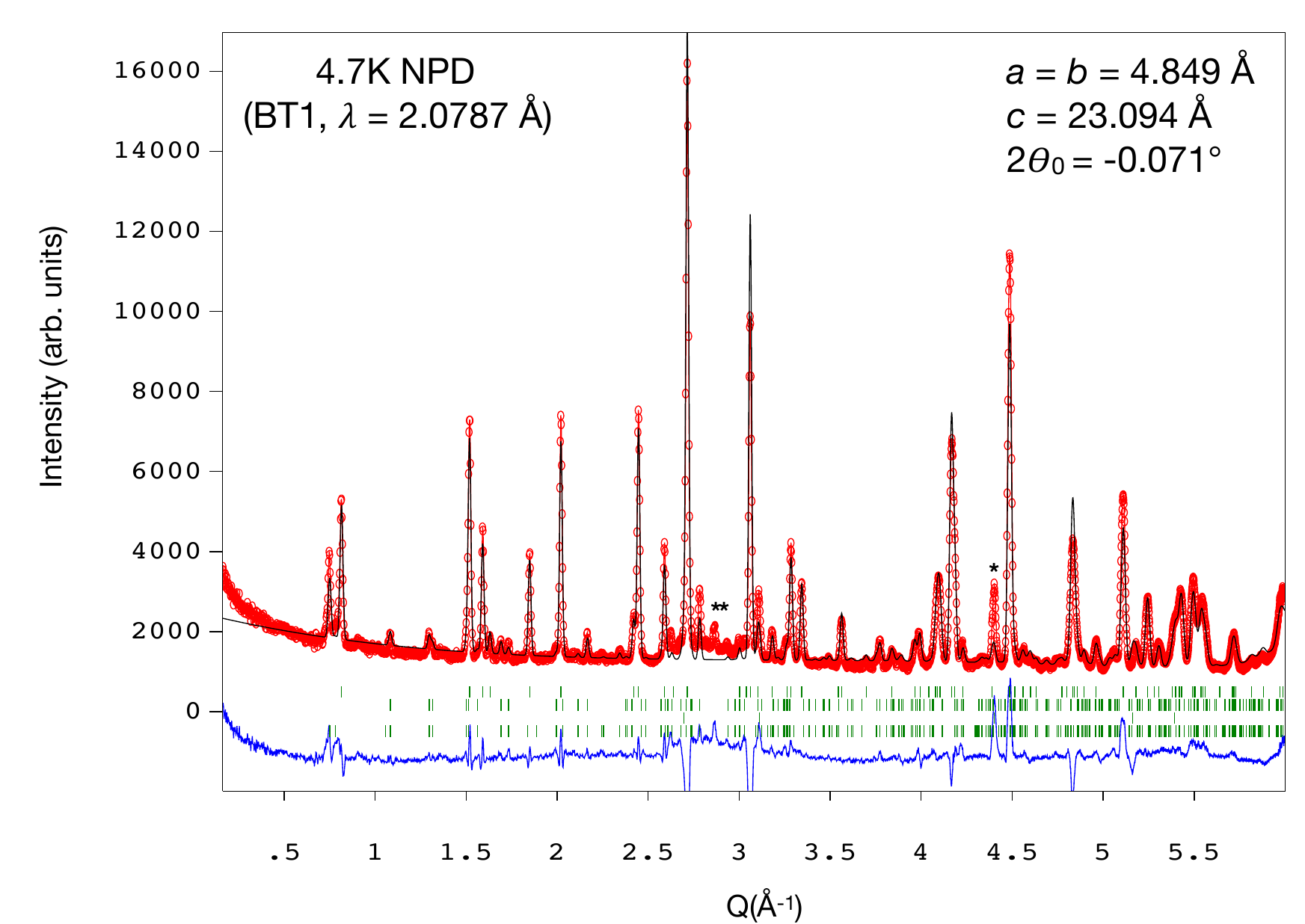}
\caption{Refinement of the nuclear structure of sample C from neutron powder diffraction (NPD) measurements on BT1.  The phases are, in order from top to bottom tick marks, \gbcpo\ (nuclear structure), Co$_2$(OH)(PO$_4$) (nuclear structure), and Co$_2$(OH)(PO$_4$) (magnetic structure).  Peaks indicated by asterisks are due to sample environment (aluminum windows on cryostat, not located at scattering center of diffractometer). } 
\label{fig:4p7K}
\end{figure}

A refinement of the $T= $4.7 K NPD data from BT1, without any subtraction, is shown in Figure \ref{fig:4p7K}. Three phases were used;  \gbcpo\ (nuclear structure), Co$_2$(OH)(PO$_4$) (nuclear structure), and Co$_2$(OH)(PO$_4$) (magnetic structure from Ref. \onlinecite{rojo2002spin}).  The refined low temperature lattice parameters for \gbcpo\ and the refined 2$\theta_0$ (zero offset in 2$\theta$) were used to calculate the expected $Q_0$ for $\vec{k_h} = (0.25,0,0)$, which is 0.374 \AA$^{-1}$.   This matches $Q_0$ extracted from the warren lineshape analysis above.

\section{Ordering wavevectors in phase III}
 \label{sec:LT}
\begin{figure}[!h]
\includegraphics[scale=0.5]{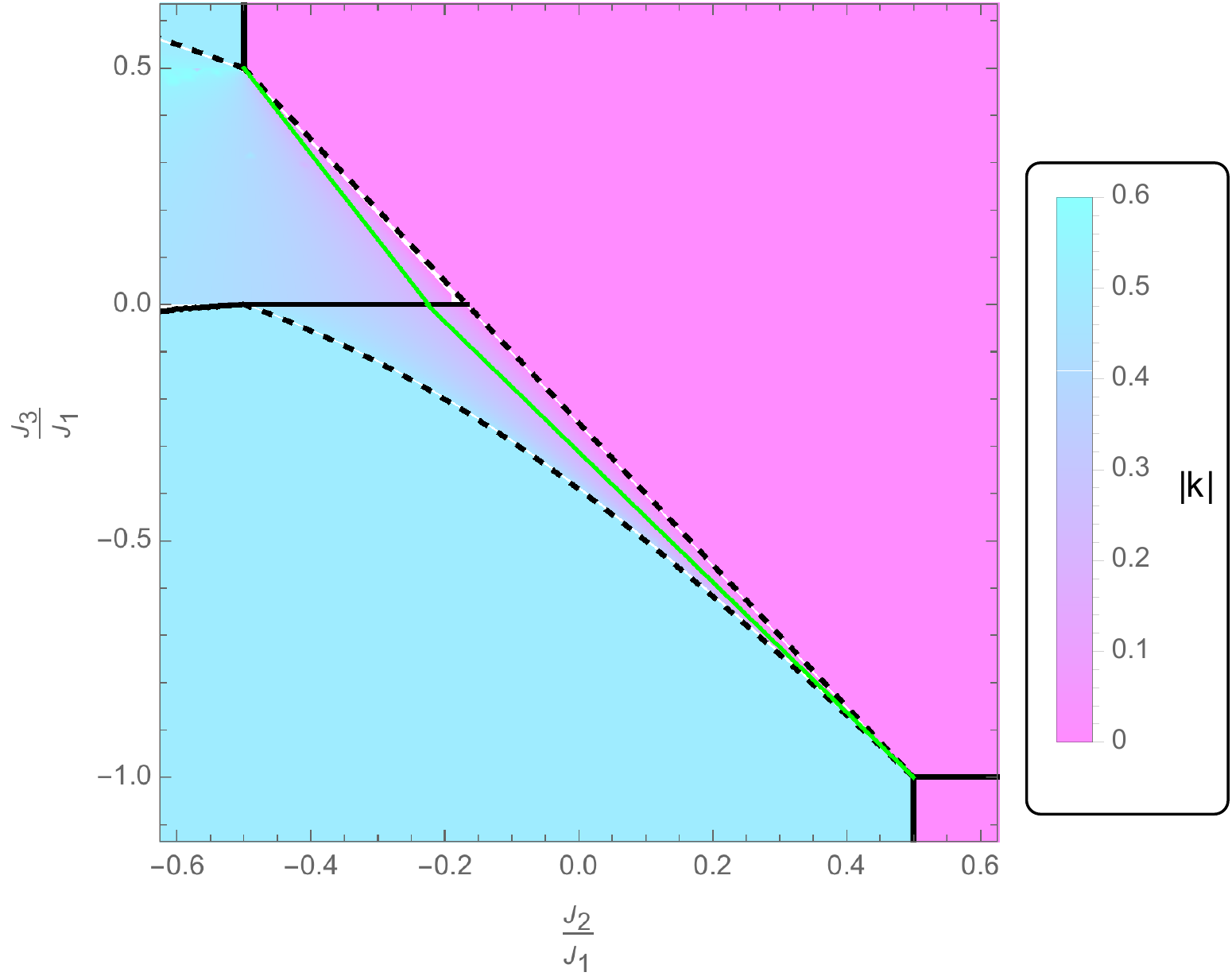}
\caption{Map of the magnitude of ordering wavevectors (reciprocal lattice units) obtained from Luttinger-Tisza analysis of the 2D honeycomb lattice \J\ Heisenberg model.  Compare to Figure 1 b in the main text for phase identification.  The green line indicates $|\vec{k}|$ = 0.25.} 

\label{fig:LT}
\end{figure}

Assuming planar spiral spin structures, Ref.~\onlinecite{rastelli1979non} obtained analytical expressions for the ordering wavevectors of the Honeycomb lattice \J\ Heisenberg model throughout the phase diagram.  However, the printed expression for phase III in Ref.~\onlinecite{rastelli1979non} does not produce real valued ordering wavevectors for all parameters in phase III.  We repeated the calculation using the modified Luttinger-Tisza method (following Ref.~\onlinecite{lyons1960method}), and the results are shown in Figure \ref{fig:LT}.  From this calculation we determined the regions of phase III and phase V which have an ordering wavevector magnitude of $|\vec{k}| = 0.25$, indicated by a green line (red dotted line in Figure 1 b in the main text).  Our results for phase V agree with the published from Ref.~\onlinecite{rastelli1979non}.

\section{Inelastic Neutron Scattering data}
\label{sec:INS}

\begin{figure*}[!t]
\includegraphics[scale=0.5]{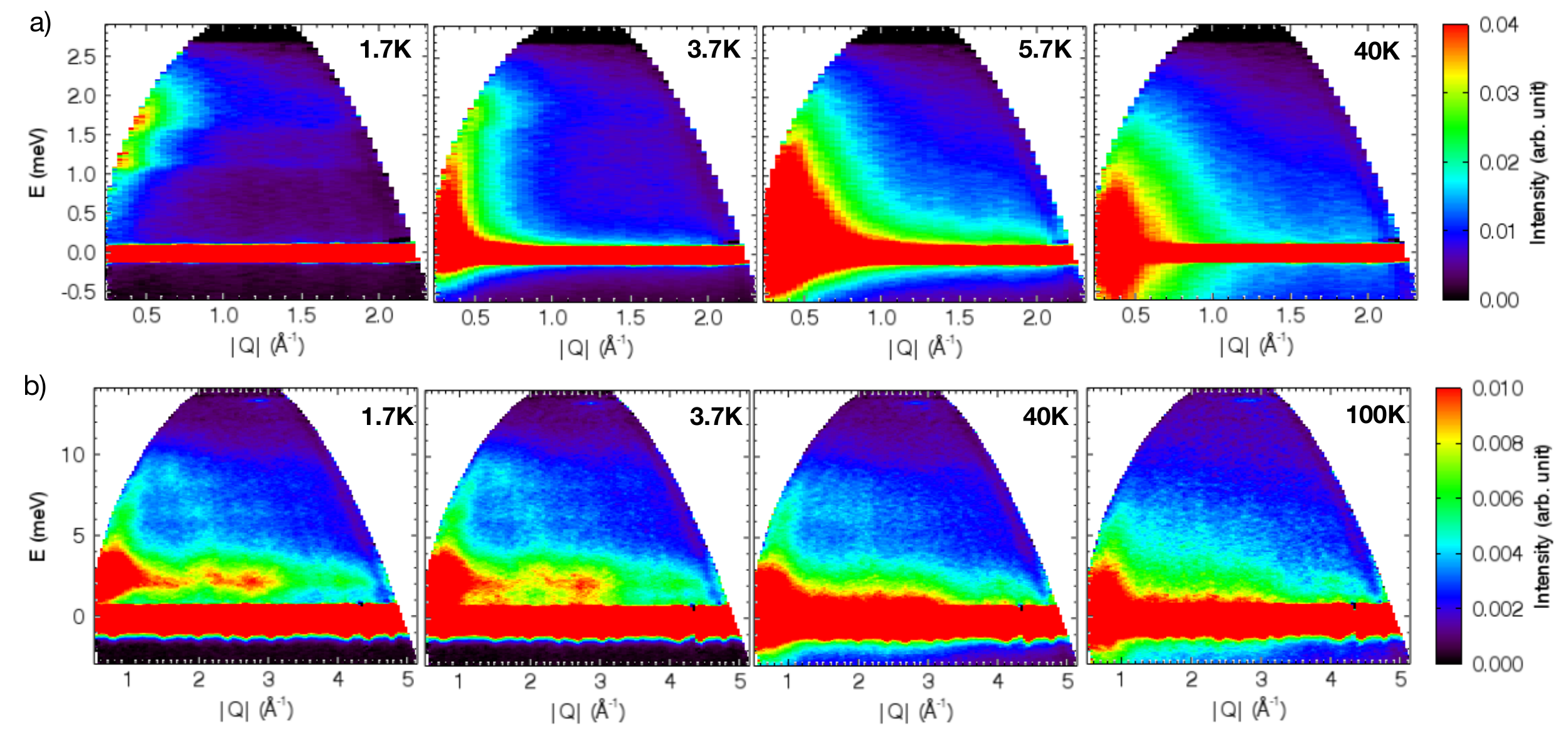}
\caption{Color contour plots of the inelastic intensities for different temperatures (empty can subtracted), taken at two different incident energies ($E_i$). a) $E_i$ = 3.07 meV, b) $E_i$ = 14.9 meV.  } 
\label{fig:INS}
\end{figure*}

Figure \ref{fig:INS} shows the full color contour maps of inelastic neutron scattering data from \gbcpo\ at selected temperatures.

\begin{figure}[!t]
\includegraphics[scale=0.3]{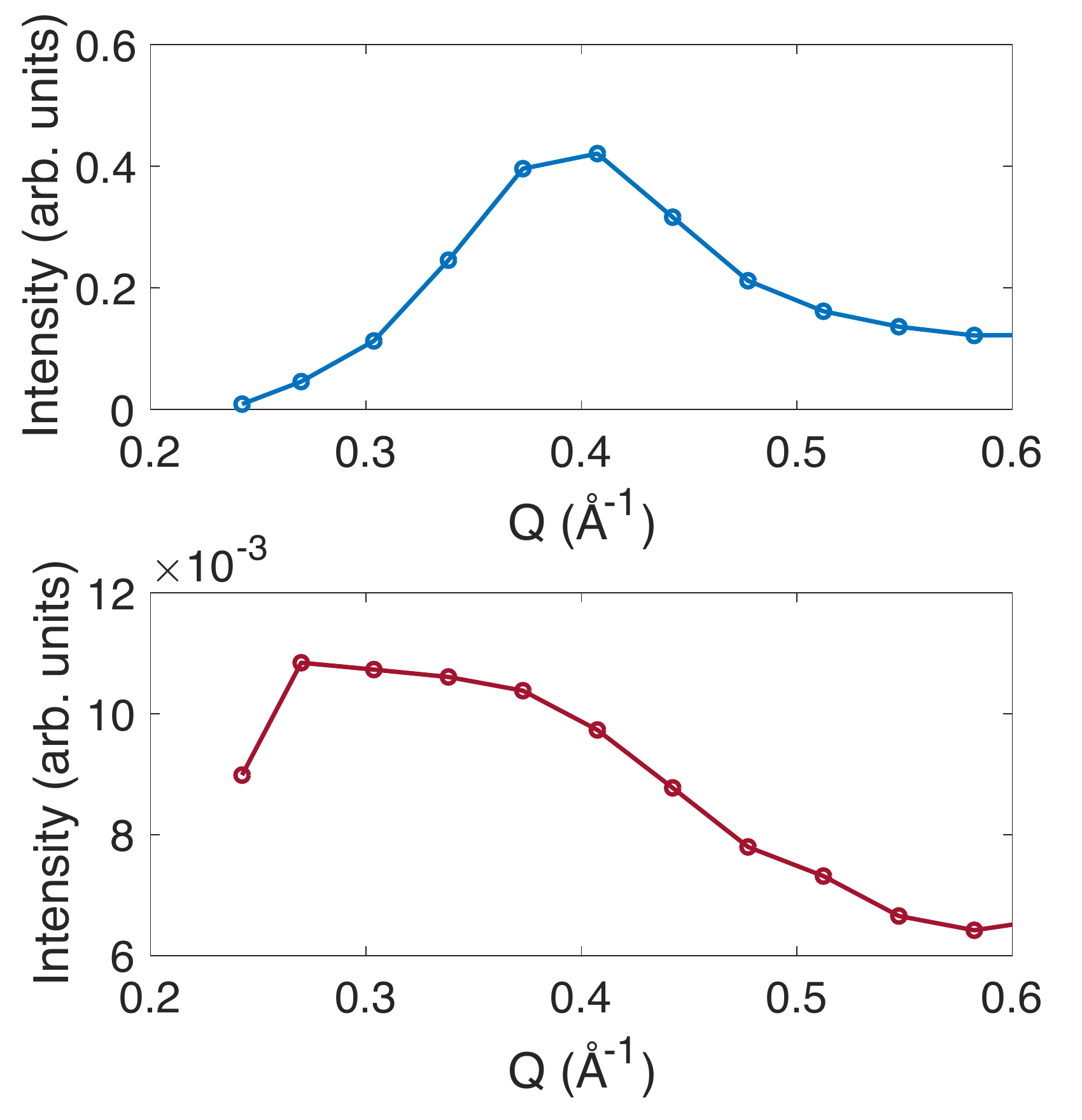}
\caption{Constant $E$ cuts at $T=1.7$K.  Top: elastic line ($E = [-0.1,0.1]$ meV), bottom: low energy inelastic ($E = [0.15, 0.2]$ meV).} 
\label{fig:INScuts}
\end{figure}

Figure \ref{fig:INScuts} shows constant energy cuts of the INS data at $T=1.7$K, for both the elastic line (-0.1 to 0.1 meV) and the low energy gapless excitation (0.15 to 0.2 meV).  The $Q$ range shown covers the lowest angle magnetic peak (from the helical short range ordered structure).   The low energy inelastic data shows a broad peak near the same wavevector, rather than being gapped and concentrated near $Q=0$ (as it is in BCAO). 

\section{Further details of Linear Spin Wave Theory results}
\label{sec:LSWT}

Here we further motivate the choice of parameter sets (1) and (2) discussed in the main text, and shown in Table \ref{tab:params}.
\begin{table}[h]
    \begin{tabular}{|c|c|c|} \hline
parameter & set 1 (phase V) & set 2 (phase III) \\ \hline
$J_1$ (meV) & -4.33 & -4.27 \\ 
$J_2$ (meV) & 0.54 & 1.92 \\ 
$J_3$ (meV) & 0.67 & -1.75 \\ 
$\lambda$ & 0.85 & 0.4 \\ 
$^{*}J_\text{perp}$ (meV) & -0.043 & -0.043 \\ \hline
$J_2/J_1$ & -0.125 & -0.449 \\ 
$J_3/J_1$ & -0.155 & 0.409 \\ \hline
\end{tabular}
\caption{Parameter sets 1 and 2.  $^{*}$Note that the results are not very sensitive to the value of $J_\text{perp}$ for $J_\text{perp} < 0.5 J_1$.   We choose a FM $J_\text{perp}$ to ensure that $k_z = 0$, so that the lowest angle magnetic reflection observed corresponds to an HK0 peak, as expected based on peak shape analysis. \label{tab:params}}
\end{table}

\begin{figure}[!h]
\includegraphics[scale=0.32]{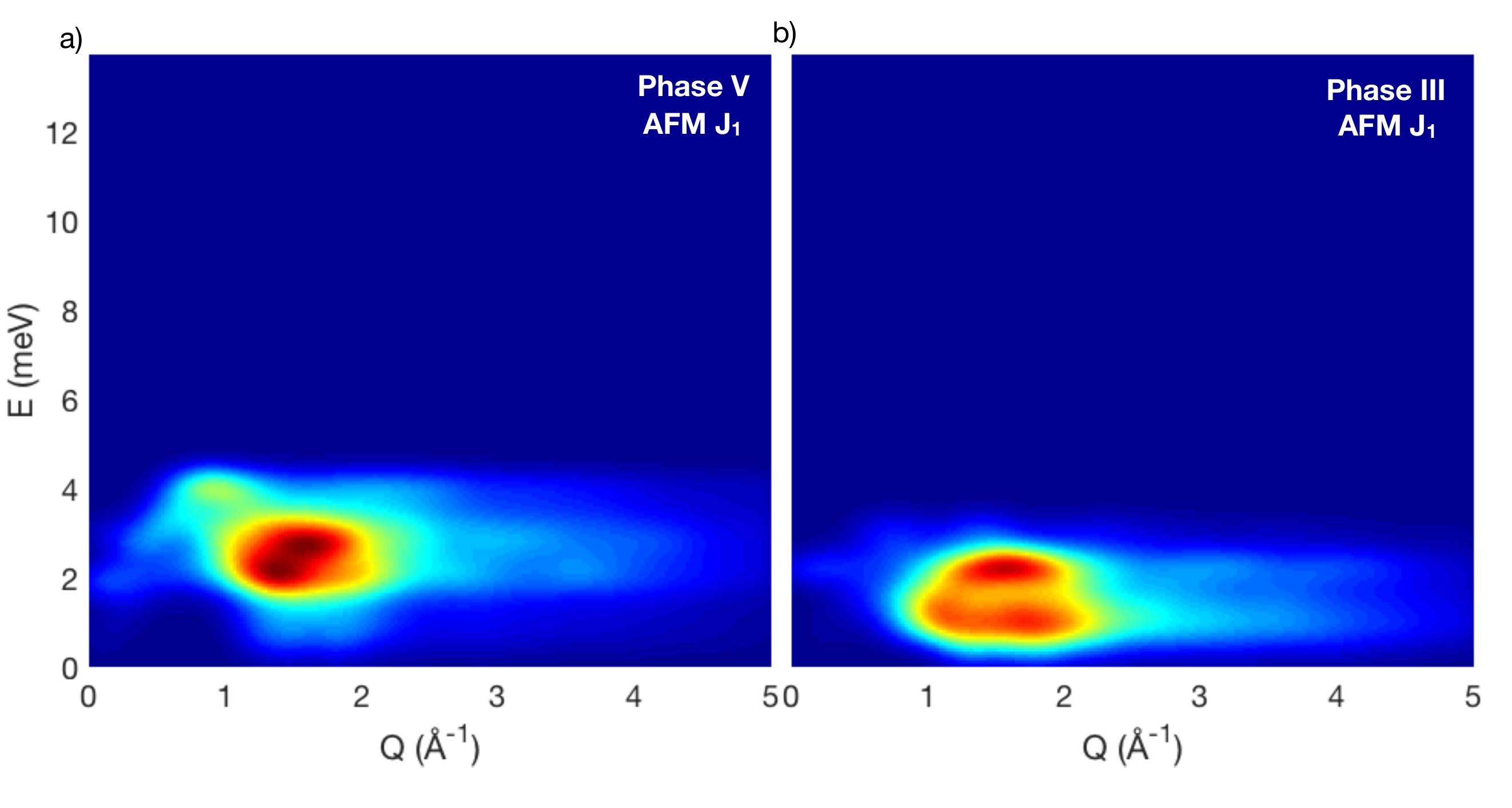}
\caption{Representative calculated spin wave spectra for $J_1>0$ (AFM near neighbor exchange), in the equivalent helical phases (phase V and phase III). The calculations were done with the equivalent parameter sets to the FM exchange sets as in Table \ref{tab:params}, i.e. $J_1 \rightarrow -J_1, J_2 \rightarrow -J_2, J_3 \rightarrow J_3$.  The AFM near neighbor model does not reproduce the main qualitative features of the spin wave spectrum, including increasing intensity as Q decreases, as well as the high energy branch visible in Figure 4 of the main text.} 
\label{fig:AFMspinwaves}
\end{figure}

 The two parameters sets are within the $J_1<0$ (FM near neighbor) model, which is motivated by the form of the inelastic intensity.  The observed spin excitation spectrum of \gbcpo\ is incompatible with AFM near neighbor interactions, which produce spectra lacking the observed intensity increase at low $Q$, as well as the high energy branch observed in the data (Figure \ref{fig:AFMspinwaves}).  FM near neighbor exchange is also anticipated based on diffuse scattering concentrated near $Q=0$ above $T_{N1}$ (Figure 3 inset in the main text).

\begin{figure}[!tb]
\includegraphics[scale=0.4]{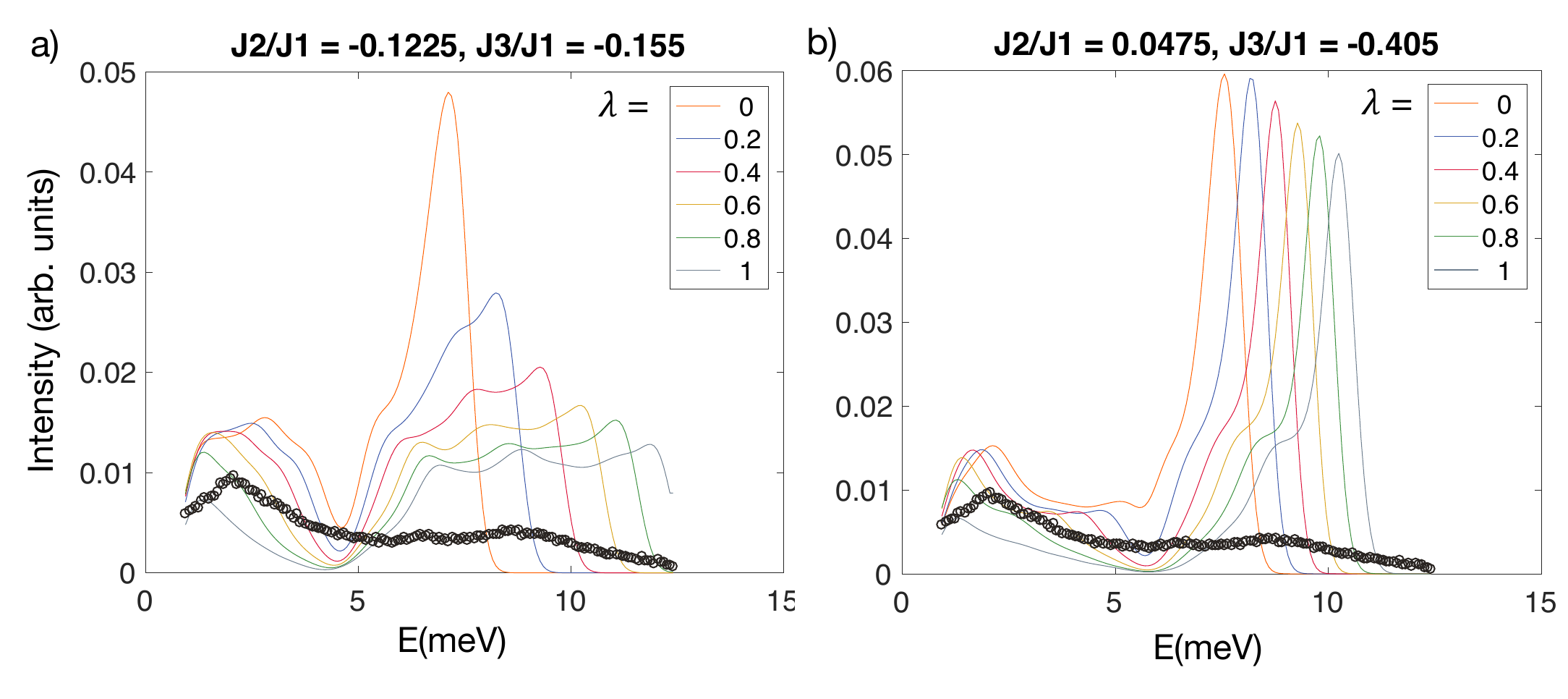}
\caption{Energy cuts integrating from $Q = 1.3$ to $1.7$ A$^{-1}$.  Comparison of data (black circles) to calculation (lines) for various $\lambda$ values, at two different points along the $|\vec{k_h}|$ = 0.25 line in phase V.  a) Parameter set 1, but with various choices of $\lambda$.  b) A point lower in phase V; note how the high energy mode remains the dominant contribution for all values of lambda for these parameters, in contrast to the data.  This occurs for increasingly negative $J_3/J_1$ values.} 
\label{fig:constQcuts}
\end{figure}

Figure \ref{fig:constQcuts} shows intensity vs. energy cuts (integrating over $Q = $1.3 to 1.7 \AA$^{-1}$) for increasingly negative $J_3$/$J_1$ values in phase V along the $\vec{k_h} = (0.25, 0, 0)$.  The intensity of the higher energy branch of excitations becomes much stronger than the lower branch, which is incompatible with our observations, and thus this parameter regime was ruled out.  The parameter sets 1 and 2 were selected based on the location of the lowest flat band (1.2 meV), the top of the high energy band (10 meV), and the relative intensity of the upper and lower bands, as discussed in the main text.

%
%

%

\end{document}